\documentclass{amsart}
\usepackage{geometry}
\usepackage{cite}
\usepackage{graphicx}
\usepackage{epstopdf}

\theoremstyle{definition}

\usepackage{enumerate}
\usepackage{multirow}
\begin{document}

\title[]{The infinite square well with a point interaction:\\
A discussion on the different parametrizations.}


\author[M. Gadella]{Manuel Gadella  }

\address{Department of Theoretical, Atomic Physics and Optics.
Facultad de Ciencias. University of Valladolid,
47011 Valladolid, Spain}
\email{manuelgadella1@gmail.com }

\author[M.A. Garc\'ia-Ferrero ]{M\textordfeminine  \'Angeles Garc\'ia-Ferrero }

\address{Department of Theoretical, Atomic Physics and Optics.
Facultad de Ciencias. University of Valladolid,
47011 Valladolid, Spain}
\email{mariangelesgferrero@gmail.com }

\author[S. Gonz\'alez-Mart\'in]{Sergio Gonz\'alez-Mart\'in}

\address{Department of Theoretical, Atomic Physics and Optics.
Facultad de Ciencias. University of Valladolid,
47011 Valladolid, Spain}
\email{sergio.gonzalez.martin@csic.es }

\author[F.H. Maldonado-Villamizar]{F\'elix H. Maldonado-Villamizar}

\address{Departamento de F\'isica. Centro de
Investigaci\'on y Estudios Avanzados del IPN. 07360, M\'exico DF.
M\'exico}
\email{felixmaldonadov@gmail.com}

\begin{abstract}
The construction of Dirac delta type potentials has been achieved
with the use of the theory of self adjoint extensions of non-self
adjoint formally Hermitian (symmetric) operators. The application of
this formalism to investigate the possible self adjoint extensions
of the one dimensional kinematic operator $K=-d^2/dx^2$ on the
infinite square well potential is quite illustrative and has been
given elsewhere. This requires the definition and use of four
independent real parameters, which relate the boundary values of the
wave functions at the walls. By means of a different approach, that
fixes matching conditions at the origin for the wave functions, it
is possible to define a perturbation of the type
$a\delta(x)+b\delta'(x)$, thus depending on two parameters, on the
infinite square well. The objective of this paper is to investigate
whether these two approaches are compatible in the sense that
perturbations like $a\delta(x)+b\delta'(x)$ can be fixed and
determined using the first approach. 
\end{abstract}

\maketitle

\section{Introduction}
\label{intro}

The question on whether a formally Hermitian operator is or it is
not self adjoint has been completely solved by mathematicians
decades ago. A very interesting presentation of this question from
the physicists point of view is given in a paper by Bonneau, Faraut
and Valent \cite{BFV}. This paper discusses the notion of self
adjoint extensions of the one dimensional momentum operator $p=-i
\frac{d}{dx}$ and the one dimensional kinetic operator
$K=-\frac{d^2}{dx^2}$ on the Hilbert space of square integrable
functions on a bounded interval (we take $m=1/2$ and $\hbar=1$ along
this Introduction and most of Section 2 for simplicity.
Nevertheless, we shall reintroduce explicitly the mass in the final
discussion).  It is shown the existence of infinite self adjoint
realizations of these operators each realization corresponding to
one distinct self adjoint operator and therefore, according to the
widely accepted interpretation, to one distinct quantum observable.

The crucial point resides in the fact that both $p=-i \frac{d}{dx}$
and $K=-\frac{d^2}{dx^2}$ belong to a special type of operators on
Hilbert space, the closed unbounded operators. Most of observables
(position, momentum, components of the angular momentum, most of
Hamiltonians) are represented by self adjoint unbounded operators.
Unbounded operators are not defined in general on the whole Hilbert
space $\mathcal H$, but on a dense subspace of $\mathcal H$, the
{\it domain} of the operator. An unbounded operator $A$ on $\mathcal
H$ is determined by both its domain ${\mathcal D}_A$ and the action
of $A$ on each $\psi\in{\mathcal D}_A$, which is $A\psi$.

Let us go back to the operator $K=-\frac{d^2}{dx^2}$ this time as an
operator on the Hilbert space $L^2([-c,c])$. This operator cannot be
defined on the whole $L^2([-c,c])$ as we know the existence of
functions on this space which either are not differentiable or do
not have square integrable derivatives. In addition, even for
domains such that both conditions are satisfied, $K$ may not be even
Hermitian, i.e., $\langle
K\psi|\varphi\rangle=\langle\psi|K\varphi\rangle$ for any pair of
functions $\psi,\varphi$ in the domain. Furthermore, Hermiticity
does not imply self adjointness, when we deal with unbounded
operators. In any case, $K$ has an infinite number of self adjoint
determinations each one characterized by its own domain. The action
of all these determinations on a given function always transform
this function into minus its second derivative, but the spaces of
functions on which they act are different.

In order to find the self adjoint determinations of $K$ one uses the
theory of extensions\footnote{The operator $B$ extends the operator
$A$ if ${\mathcal D}_A\subset{\mathcal D}_B$ and $A\psi=B\psi$ for
all $\psi\in{\mathcal D}_A$. Then, we write $A\prec B$.} of
Hermitian operators. It is not our intention to give here a review
of this theory. For a presentation comprehensible to physicists, see
\cite{BFV}.

For the infinite square well, we have chosen a point perturbation of
the type $a\delta(x)+b\delta'(x)$, where $\delta(x)$ is the Dirac
delta and $\delta'(x)$ its derivative in the distributional sense,
on an infinite square well centered at the origin. In this case, the
formal Hamiltonian takes the form:

\begin{equation}\label{1}
H= -\frac1{2m}\,\frac{d^2 }{dx^2}+V(x)+a\delta(x)+b\delta'(x)
\end{equation}
with

\begin{equation}\label{2}
\quad V(x)=\left\{ \begin{array}{ccc}
\infty  & {\rm if} &  x<-c \\
0 & {\rm if} & -c\le x\le c \\
\infty  & {\rm if} &  x>c       \end{array}
  \right.\,.
\end{equation}

In consequence, the time independent Schr\"odinger equation is

\begin{equation}\label{3}
-\frac1{2m}\,\frac{d^2
f(x)}{dx^2}+\{V(x)+a\delta(x)+b\delta'(x)\}f(x)=Ef(x)\,,
\end{equation}
which is an equation on distributions.

Functions in the domain for which the Hamiltonian in (\ref{3}) is
self adjoint cannot be continuous and with continuous derivative at
the origin \cite{K,GO,Z}. Therefore, we need to define the
distributions resulting from the products of $\delta(x)$ and
$\delta'(x)$ times one function discontinuous at the origin.
Henceforth, we shall the following definitions:

\begin{eqnarray}
&&f(x) \delta(x)=\frac{f(0-)+f(0+)}{2}\,\delta(x)\,, \label{4}\\[2ex]
&&f(x) \delta'(x)=
\frac{f(0-)+f(0+)}{2}\,\delta'(x)-\frac{f'(0-)+f'(0+)}{2}\,\delta(x)\,,\label{5}
\end{eqnarray}
where,

\begin{equation}\label{6}
f(0-)=\lim_{x\mapsto 0^-} f(x)\,,\qquad f(0+)=\lim_{x\mapsto 0^+}
f(x)\,.
\end{equation}
Same for $f'(0-)$ and $f'(0+)$.

Concerning the derivative of the delta: It is not well known that
the term in the form $b\delta'(x)$ does not provide a unique
perturbation of the Hamiltonian
$-\frac1{2m}\,\frac{d^2}{dx^2}+V(x)+a\delta(x)$. In fact, the
introduction of the term $b\delta'(x)$ often produces a certain
degree of confusion. For instance, some authors say that if we add
to a potential a term of this kind, then the potential is opaque,
i.e., no transmission coefficient exists. On the other hand, other
authors find a non-zero transmission coefficient for
$a\delta(x)+b\delta'(x)$. The reason of this disagreement lies on
the use of different self adjoint extensions that provide different
realizations for the perturbation $b\delta'(x)$. See
\cite{SE,TN,ZZ,GO,Z}. In this paper, we shall define a self adjoint
realization of (\ref{1}) having reasonable physical properties such
as non-zero transmission and reflection coefficients through the
$a\delta(x)+b\delta'(x)$ barrier, as in \cite{GNN}.

Self adjoint extensions of the kinetic operator $K=-d^2/dx^2$ on the
infinite square well have been discussed in \cite{BFV}. These self
adjoint extensions are parameterized by five real numbers having one
relation among them, so that only four are independent. This means
that each of the self adjoint extensions of $K$ is characterized by
the actual values of four real parameters. These parameters relate
the boundary values of the wave functions and of their first
derivatives as given at both walls of the well.

Our Hamiltonian $H$ in (\ref{1}) is given by a point perturbation
added to $K$ depending on two real parameters. This perturbation is
obtained by choosing a suitable self adjoint extension of $K$.
However, in this case, this self adjoint extension is determined by
some matching conditions imposed to the wave functions at the
origin. Then, the question that we want to investigate here is how
we could characterize this self adjoint extension using the
parameters relating the boundary conditions at the wall as discussed
in \cite{BFV}. As we shall see, this is not a trivial matter.

This paper is organized as follows: In Section 2, we introduce two
possible parameterizations of self adjoint extensions of the kinetic
operator in a finite interval, just the parameterizations we want to
compare.  Section 3 contains the core of the present work. We use
the results in \cite{BFV} to determine the parameters that produce
the self adjoint determination of (\ref{1}) that we are considering.
The final conclusion shows that the relation between parameters is
not one to one.

\section{The infinite square well with a point perturbation}
\label{sec:1}

Let us consider the Hamiltonian of a one dimensional free particle
confined in the interval $[-c,c]$. Its Hamiltonian is given by
$H=-d^2/dx^2+V(x)$, where $V(x)$ is given by (\ref{2}), the infinite
square well potential. Along the present section, we shall usually
take $m=1/2$ for simplicity, although we shall explicitly show the
mass $m$ whenever convenient.

As a matter of fact, the issue here is the analysis of the
differential operator $K=-d^2/dx^2$, sometimes called the kinetic
operator, on the interval $[-c,c]$. As is well known, this is an
unbounded operator which is not completely determined until we
define its domain, i.e., the space of vectors on which it acts. The
Hilbert space of pure states for this interval is $L^2[-c,c]$.
Therefore, the domain of $K=-d^2/dx^2$ should be contained in the
space of square integrable functions on $[-c,c]$ which are twice
differentiable (almost elsewhere, i.e. with the possible exception
of points in a set of zero Lebesgue measure) and such that their
first and second derivative are also square integrable on the same
interval. We shall call this space\footnote{Technically, ${\mathcal
D}^*$ is the space of absolutely continuous functions in $L^2[-c,c]$
with first absolutely continuous derivative and such that
$\int_{-c}^c \{|f(x)|^2+|f''(x)|^2\}\,dx<\infty$.} ${\mathcal D}^*$.

Now, the question is to find out the domains for which this operator
is self adjoint. Let us consider the functions $\psi(x)$ and
$\varphi(x)$ in ${\mathcal D}^*$. Then, integration by parts gives:


\begin{equation}
\langle\psi|-\frac{d^2}{dx^2}\,\varphi\rangle = -i\{
\psi^*(c)\varphi'(c)-\psi^*(-c)\varphi'(-c)+{\psi'}^*(c)\varphi(c)-{\psi'}^*(-c)\varphi(-c)\}
+\langle-\frac{d^2}{dx^2}\,\psi|\varphi\rangle\,.
\label{7}
\end{equation}
In (\ref{7}), the star denotes complex conjugation. Note that we
demand the square integrability of the first derivatives of
functions in ${\mathcal D}^*$ in order to be able to integrate by
parts.

Obviously, $K=-d^2/dx^2$ on ${\mathcal D}^*$ is not Hermitian. By
Hermiticity, we mean that $\langle\varphi|K\psi\rangle=\langle
K\varphi|\psi\rangle$ for any pair of functions
$\varphi\equiv\varphi(x)\,,\psi\equiv \psi(x)\in {\mathcal D}^*$. We
note  that the necessary and sufficient condition for the
Hermiticity of $K=-d^2/dx^2$ is that

\begin{equation}\label{8}
\psi^*(c)\varphi'(c)-\psi^*(-c)\varphi'(-c)+{\psi'}^*(c)\varphi(c)-{\psi'}^*(-c)\varphi(-c)=0\,.
\end{equation}

We need to choose a domain ${\mathcal D}$ for $K=-d^2/dx^2$ with the
obvious condition that ${\mathcal D}\subset {\mathcal D}^*$, also
fulfilling (\ref{8}). In order to define ${\mathcal D}$ one may
choose all functions $\varphi(x)\in{\mathcal D}^*$ such that
$\varphi(-c)=\varphi(c)=\varphi'(-c)=\varphi'(c)=0$. Let us denote
by ${\mathcal D}_0$ the space of the functions in ${\mathcal D}^*$
with this property. Any function $\varphi(x)\in{\mathcal D}_0$ and
its first derivative are continuous at the well borders $c$ and
$-c$. Clearly, $K=-d^2/dx^2$ is Hermitian in ${\mathcal D}_0$.
However, $K$ is not self adjoint. Due to the Hermiticity of $K$,
$K\prec K^\dagger$, i.e., the adjoint of $K$, $K^\dagger$ extends
$K$. One can easily prove that this extension is strict, so that
$K\ne K^\dagger$ and therefore $K$ cannot be self adjoint.

The search for self adjoint extensions of $K$ is nothing else that
the search for domains $\mathcal D$ for $K$ with the condition that
any $\psi(x)\in\mathcal D$ satisfies (\ref{8}) without making use of
the trivial condition $\psi(-c)=\psi(c)=\psi'(-c)=\psi'(c)=0$. Then,
the domains of $K$ and $K^\dagger$ will coincide and therefore
$K=K^\dagger$. This problem has been solved and extensively
discussed in the literature \cite{BFV,K}. The domains that make the
operator self adjoint are the spaces of functions
$\psi(x)\in{\mathcal D}^*\subset L^2[-c,c]$ such that

\begin{equation}\label{9}
\left(
  \begin{array}{c}
 2c\psi'(-c)-i\psi(-c) \\[2ex]
    2c\psi'(c)+i\psi(c) \\
  \end{array}
\right)
 ={\bf U} \left(
            \begin{array}{c}
 2c\psi'(-c)+i\psi(-c) \\[2ex]
2c\psi'(c)-i\psi(c) \\
            \end{array}
          \right)\,,
\end{equation}
where the 
matrix $\bf U$ depends on four real parameters and has the form
\cite{BFV}

\begin{equation}\label{10}
{\bf U}=e^{i\Phi}\left(
                   \begin{array}{cc}
m_0-im_3 & -m_2-im_1 \\[2ex]
m_2-im_1 & m_0+im_3 \\
                   \end{array}
                 \right)\,,
\end{equation}
with
\begin{equation}\label{11}
m_0^2+m_1^2+m_2^2+m_3^2=1\qquad {\rm and} \qquad \Phi\in[0,\pi]\,.
\end{equation}

We see that there are five parameters and one relation between them,
so that there are indeed four independent parameters. Each set of
fixed values of these parameters gives a self adjoint extension of
$K=-d^2/dx^2$. Each of the self adjoint extensions is a different
operator with purely discrete spectrum.

As we have already remarked, there is another possibility, another
characterization of the domains of the different self adjoint
extensions of $K=-d^2/dx^2$, which consists in fixing matching
conditions at the origin in the spirit of \cite{K}. We would like to
know how the correspondence between these two approaches are and
particularly how the correspondence between the parameters that
label the self adjoint extensions of $K$ looks like. In general,
this seems a rather cumbersome task. We shall limit our analysis to
the case of the extensions of $K$ producing the $a\delta+b\delta'$
perturbation in the Hamiltonian $H=-\frac1{2m}\,d^2/dx^2+V(x)$ with
$V(x)$ as in (\ref{2}). This will give enough relevant information
on how this kind of correspondence between the parameters defining
the self adjoint determination (extension) in two different settings
work. In addition, the problem in its full generality is not
tractable. It is important to remark that two different self adjoint
extensions of $K$ are different operators and have different
eigenvalues \cite{BFV}.

\subsection{The one dimensional infinite square well with a point
perturbation of the type \break $\mathbf{a\boldsymbol\delta(x)\boldsymbol+b\boldsymbol\delta'(x)}$} \label{sec:2}

Next let us make a brief excursion into the operator $-d^2/dx^2$ on
$L^2({\mathbb R})$.  One possible domain, ${\mathcal D}_{0,\infty}$,
for $-d^2/dx^2$ is the vector space of square functions in
$L^2({\mathbb R})$ such that: i.) admit a first and second
derivative which are square integrable (indeed it suffices that the
derivative exists save for a null set, but we ignore here certain
mathematical technicalities), ii.) so that all functions $\psi(x)$
in ${\mathcal D}_{0,\infty}$ satisfy $\psi(0)=\psi'(0)=0$ at the
origin and iii.) at the infinity we have
$\psi(-\infty)=\psi(\infty)=0$\footnote{Or in more technical terms,
the Sobolev space $W^2_2({\mathbb R})$.}. Here, we want to remark
that although a square integrable function may not have a limit at
the infinity\footnote{This function may be even of class $C^\infty$ on the whole real line. See an example in the Appendix of \cite{AGW}}, if this limit exists it must be zero.

In this case, it is a simple exercise to see that the domain of the
adjoint is the space of functions in $L^2({\mathbb R})$ satisfying
i.) and iii.), with ii.) replaced by the condition that both
$\psi(x)$ and its derivative $\psi'(x)$ have a finite discontinuity
or jump at the origin\footnote{Here, we have avoided some
technicalities. As a matter of fact this domain is the Sobolev space
$W^2_2({\mathbb R}\setminus\{0\})$ \cite{K}.}. On this domain, the adjoint
acts exactly as $-d^2/dx^2$ does.

Now, let us assume that $\varphi(x)$ and $\psi(x)$ belong to the
domain of the adjoint $(-d^2/dx^2)^\dagger$ of $-d^2/dx^2$. Then, if
we denote the left and right limits at the origin of a function
$\phi(x)$ by $\phi(0-)$ and $\phi(0+)$ respectively (as in
(\ref{6})), we have by integration by parts:

\begin{eqnarray*}
 \langle \varphi|\left(-\frac{d^2}{dx^2} \right)^\dagger
 \psi\rangle &=& 
 -\int_{-\infty}^\infty
\varphi(x)\psi''(x)\,dx= = -\int_{-\infty}^0 \varphi(x)\psi''(x)\,dx
-\int_{0}^\infty \varphi(x)\psi''(x)\,dx=\nonumber\\[2ex] 
&=&-\{\varphi(0-)\psi'(0-)-\varphi(-\infty)\psi'(-\infty)\}
-\{\varphi(\infty)\psi'(\infty)-\varphi(0+)\psi'(0+)\}+
\\[2ex] && +
\{\varphi'(0-)\psi(0-)-\varphi'(-\infty)\psi(-\infty)\} +
\{\varphi'(\infty)\psi(\infty)-\varphi'(0+)\psi(0+)\}-
\nonumber\\[2ex]&& -\int_{-\infty}^0 \varphi''(x)\psi(x)\,dx
-\int_{0}^\infty \varphi''(x)\psi(x)\,dx.
\end{eqnarray*}

Taken into account that the functions in the domain of the adjoint
vanish at the infinity, the above expression is equal to

\begin{eqnarray}
-\{\varphi(0-)\psi'(0-)-\varphi(0+)\psi'(0+)\}+
\{\varphi'(0-)\psi(0-)-\varphi'(0+)\psi(0+)\} 
-\int_{-\infty}^\infty \varphi''(x)\psi(x)\,dx=\nonumber\\[2ex]=\langle
\left(-\frac{d^2}{dx^2} \right)^\dagger\varphi|\psi\rangle\,.
\label{12}
\end{eqnarray}

As in the previous discussion about the operator $K=-d^2/dx^2$ on
the infinite square well, in order to obtain the self adjoint
extensions of this operator, we have to find the spaces of functions
for which (\ref{12}) vanishes identically, excluding the trivial
possibility given by (\ref{8}). Then, these self adjoint extensions
will be determined by $-d^2/dx^2$ operating on each of these
domains.

Each one of these self adjoint extensions is characterized by the
fact that their functions $\psi(x)$ satisfy relations of the type
\cite{K}:

\begin{equation}\label{14}
\left(
  \begin{array}{c}
\psi(0+) \\[2ex]
\psi'(0+) \\
  \end{array}
\right)= \left(
  \begin{array}{cc}
\frac{(2+x_2)^2-x_1x_4+x_3^2}{(2-ix_3)^2+x_1x_4-x_2^2} & \frac{-4x_4}{(2-ix_3)^2+x_1x_4-x_2^2} \\[2ex]
\frac{4x_1}{(2-ix_3)^2+x_1x_4-x_2^2} & \frac{(2-x_2)^2-x_1x_4+x_3^2}{(2-ix_3)^2+x_1x_4-x_2^2} \\
  \end{array}
\right) \left(
   \begin{array}{c}
\psi(0-) \\[2ex]
\psi'(0-) \\
   \end{array}
 \right)\,.
\end{equation}
Each set of values of the four real parameters, $x_i$, $i=1,2,3,4$,
determines one self adjoint extension of $-d^2/dx^2$ \cite{K}.
However, we are not interested here in all self adjoint extensions,
which are anyway listed in \cite{K}.

The interesting point is that we can define point potentials of the
type $a\delta(x)+b\delta'(x)$ by means of these self adjoint
extensions \cite{K,AL,GO}. This can be achieved if we choose the
following values for the parameters: $x_1=a$, $x_2=b$, $x_3=x_4=0$
\cite{K}. Note that the simplest choice, $x_1=x_2=x_3=x_4=0$,
produces the identity matrix in (\ref{14}). It also determines a
self adjoint extension of $-d^2/dx^2$.

If we recover the arbitrary value for the mass (as we shall do
consistently in the final section), we may write the Hamiltonian
corresponding to this particular extension as
$-d^2/dx^2+2ma\delta(x)+2mb\delta'(x)$. Then, $x_1=2ma$, $x_2=2mb$
and $x_3=x_4=0$. Thus, (\ref{14}) takes the following form:

\begin{equation}\label{15}
\left(
  \begin{array}{c}
\psi(0+) \\[2ex]
\psi'(0+) \\
  \end{array}
\right)= \left(
  \begin{array}{cc}
\frac{1+mb}{1-mb} & 0 \\[2ex]
\frac{-2ma}{1-m^2b^2} & \frac{1-mb}{1+mb} \\
  \end{array}
\right) \left(
   \begin{array}{c}
\psi(0-) \\[2ex]
\psi'(0-) \\
   \end{array}
 \right)\,.
\end{equation}

Relation (\ref{15}) determines the domain of the self adjoint
extension of $-d^2/dx^2$ ($-\frac1{2m}\,\frac{d^2}{dx^2}$)
corresponding to the Hamiltonian given by
$-d^2/dx^2+2ma\delta(x)+2mb\delta'(x)$
($-\frac1{2m}\;d^2/dx^2+a\delta(x)+b\delta'(x)$).

Now let us go back to the case in which $K=-d^2/dx^2$ is defined on
the Hilbert space $L^2[-c,c]$, i.e., is the operator relative to the
infinite one dimensional square well studied in the previous
version. In order to define a perturbation of the type
$a\delta(x)+b\delta'(x)$ on the infinite square well, we still need
to define the self adjoint extension of $K$ using matching
conditions (\ref{15}). Now, the objective is to investigate how we
can obtain this perturbation starting with conditions (\ref{9}) and
(\ref{10}). This is the objective of the next section.

In a previous paper \cite{GGN}, we have discussed the effect on a
one dimensional infinite square well of a perturbation of the free
Hamiltonian of the type $a\delta(x)+b\delta'(x)$. We have analyzed
how the eigenvalues behave under changes of $a$ and $b$. We want to
compare formulas (\ref{9}) and (\ref{10}) to (\ref{15}) in order to
identify which parameters in (\ref{10}) correspond to this
perturbation. This would permit us to compare the results for the
energy levels obtained in \cite{GGN} with those in \cite{BFV}. This
is the main objective of the present work and will be developed in
the next section.

\section{Parameters of the self adjoint extension defining the perturbation
$a\delta(x)+b\delta'(x)$  centered on the infinite square well}
\label{sec:3}

This section contains the main objective of the present paper. As we
have remarked, we want to discuss the relation between the
determination of self adjoint extensions of $K=-d^2/dx^2$ given by
the boundary conditions (\ref{9}) and the matching conditions
(\ref{14}). However, this problem in its full generality seems too
difficult and even untractable, so that we shall undergo a simpler
task: the relation between (\ref{9}) and (\ref{15}). As we know,
boundary conditions (\ref{15}) determine the Hamiltonian with point
potential (\ref{1}). Therefore, our investigation consists in
finding the values of the parameters in (\ref{9}) that give the
point potential $a\delta(x)+b\delta'(x)$. As we shall see along the
next lines, this is not a particularly simple task and the final
result is not simple.

To begin with,  the solutions of the Schr\"odinger equation on the
infinite square well

$$-\psi''(x)+2ma\delta(x)\psi(x)+2mb\delta'(x)=2mE\psi(x)$$ are given
by the following plane waves:

\begin{eqnarray}
\psi_1(x) &=& De^{ikx}+Ce^{-ikx}\,, \qquad -c<x<0\,, \label{16}
\\[2ex]
\psi_2(x) &=& Ae^{ikx}+Be^{-ikx}\,, \qquad 0>x>c\,. \label{17}
              \end{eqnarray}

Note that $\psi_1(x)$ and $\psi_2(x)$ are the solutions to the left
and to the right respectively of the origin. At the origin, we
assume that t(\ref{16}) and (\ref{17}) satisfy (\ref{15}). Let us use these
results in equations (\ref{9}) and (\ref{10}). We obtain:

\begin{eqnarray}\label{18}
\left(\begin{array}{c}D\beta e^{-ikc}-C\alpha e^{ikc}\\ A\alpha
e^{ikc}-B\beta
e^{-ikc}\end{array}\right)=\left(\begin{array}{cc}U_{11}&U_{12}\\U_{21}&U_{22}\end{array}
\right)\left(\begin{array}{c}D\alpha e^{-ikc}-C\beta e^{ikc}\\
A\beta e^{ikc}-B\alpha e^{-ikc}\end{array}\right)\,,
\end{eqnarray}
where $\alpha=2ck+1$ and $\beta=2ck-1$ and $U_{ij}$ are the entries
of matrix $\bf U$ given in (\ref{10}). We write $\bf U$ in this form
just for convenience in our presentation and also in order to
simplify our calculations as much as possible. It is straightforward
that we can write (\ref{18}) as

\begin{equation}
\left(\begin{array}{c}A\beta U_{12} e^{ikc}-B\alpha U_{12}
e^{-ikc}\\[2ex] A\left(\alpha-U_{22}\beta\right)
e^{ikc}-B\left(\beta-U_{22}\alpha\right)
e^{-ikc}\end{array}\right)= 
\left(
\begin{array}{c}D\left(\beta-U_{11}\alpha\right)
e^{-ikc}-C\left(\alpha-U_{11}\beta\right) e^{ikc}\\[2ex] DU_{21}\alpha
e^{-ikc}-CU_{21}\beta e^{ikc}\end{array}\right)\,. \label{19}
\end{equation}

Equation (\ref{19}) can obviously be rewritten in abridged form as:

\begin{equation}\label{20}
\mathbf{R}\left(\begin{array}{c}A\\
B\end{array}\right)=\mathbf{V}\left(\begin{array}{c}D\\
C\end{array}\right)\,,
\end{equation}
where

\begin{eqnarray}\label{21}\nonumber
\mathbf{R}&=&\left(\begin{array}{cc}U_{12}\beta e^{ikc}&-U_{12}\alpha e^{-ikc}\\
&\\\left(\alpha-U_{22}\beta\right)e^{ikc}&-\left(\beta-U_{22}\alpha\right)e^{-ikc}
\end{array}\right)\\\nonumber\\\nonumber\\
\mathbf{V}&=&\left(\begin{array}{cc}\left(\beta-U_{11}\alpha\right)e^{-ikc}&-
\left(\alpha-U_{11}\beta\right)e^{ikc}\\[2ex]
U_{21}\alpha e^{-ikc}&-U_{21}\beta e^{ikc}\end{array}\right)\,.
\end{eqnarray}

Equation (\ref{20}) can be obviously rewritten as:

\begin{equation}\label{22}
\left(\begin{array}{c}A\\
B\end{array}\right)=\mathbf{R^{-1}V}\left(\begin{array}{c}D\\
C\end{array}\right)\,,
\end{equation}
with

\begin{eqnarray}
\mathbf{R^{-1}}&=&\frac{1}{\Delta}\left(\begin{array}{cc}-
\left(\beta-U_{22}\alpha\right)e^{-ikc}&U_{12}\alpha e^{-ikc}\\
&\\-\left(\alpha-U_{22}\beta\right)e^{ikc}& U_{12}\beta
e^{ikc}\end{array}\right) \label{23}
\end{eqnarray}
and

\begin{equation}\label{24}
\Delta= U_{12}(\alpha^2-\beta^2) =-8ck(im_1+m_2)e^{i\Phi}\,.
\end{equation}

Now, we are going to obtain a similar result by another method and
then, compare this result with the already obtained. First of all,
let us write (\ref{15}) in accordance with the notation used in
(\ref{16}-\ref{17}),  in the following form:

\begin{equation}\label{25}
\left(
  \begin{array}{c}
\psi_2(0+) \\[2ex]
\psi'_2(0+) \\
  \end{array}
\right)= \left(
  \begin{array}{cc}
t_1 & 0 \\[2ex]
t_2 & \frac1{t_1} \\
  \end{array}
\right) \left(
   \begin{array}{c}
\psi_1(0-) \\[2ex]
\psi'_1(0-) \\
   \end{array}
 \right)\,,
\end{equation}
with

\begin{equation}\label{26}
{\bf T}=\left(
  \begin{array}{cc}
t_1 & 0 \\[2ex]
t_2 & \frac1{t_1} \\
  \end{array}
\right)= \left(
  \begin{array}{cc}
\frac{1+mb}{1-mb} & 0 \\[2ex]
\frac{-2ma}{1-m^2b^2} & \frac{1-mb}{1+mb} \\
  \end{array}
\right)\,.
\end{equation}

We write the matrix $\bf T$ in the form (\ref{25}) in order to
simplify the subsequent calculations. Then, if we use
(\ref{16}-\ref{17}) in (\ref{26}), we obtain

\begin{eqnarray}
 \left( \begin{array}{c} A+B \\[2ex] ik(A-B)\end{array}\right)=\left(
\begin{array}{cc}\frac{1+mb}{1-mb}&0\\[2ex]\frac{-2ma}{1-m^2b^2}&\frac{1-mb}{1+mb}\end{array}\right)
\left( \begin{array}{c} D+C \\[2ex] ik(D-C)\end{array}\right)\,.
\label{27}
\end{eqnarray}

This equation can be written in a similar form as in (\ref{20}). A
rather straightforward calculation gives:

\begin{eqnarray}
\left(\begin{array}{c}A\\
B\end{array}\right)=\mathbf{M^{-1}TM}\left(\begin{array}{c}D\\
C\end{array}\right)\,, \label{28}
\end{eqnarray}
with

\begin{equation}\label{29}
{\bf M}= \left(
           \begin{array}{cc}
1 & 1 \\[2ex]
ik & -ik \\
           \end{array}
         \right)\qquad {\rm and} \qquad
 {\bf M}^{-1}= \frac{1}{2ik}\left(
\begin{array}{cc}ik&1\\&\\ik&-1\end{array}\right)\,.
\end{equation}

Comparing (\ref{22}) and (\ref{28}), we have:

\begin{equation}\label{30}
\mathbf{R^{-1}V}=\mathbf{M^{-1}TM}\,.
\end{equation}

The next step is to identify matrix elements in the right and left
hand sides of (\ref{30}) in order to write a system of four
equations in the four undeterminates $U_{ij}$. This system is:

\begin{eqnarray}
-\left(\beta-U_{22}\alpha\right)\left(\beta-U_{11}\alpha\right)+U_{12}
U_{21}\alpha^2&=&\frac{ik(t_1+\frac{1}{t_1})+t_2}{2ik}e^{2ikc}\Delta\,,
\label{31}
\\[2ex]
\left(\beta-U_{22}\alpha\right)\left(\alpha-U_{11}\beta\right)-U_{12}
U_{21}\alpha\beta&=&\frac{ik(t_1-\frac{1}{t_1})+t_2}{2ik}\Delta\,,\label{32}\\[2ex]
-\left(\alpha-U_{22}\beta\right)\left(\beta-U_{11}\alpha\right)+U_{12}
U_{21}\alpha\beta&=&\frac{ik(t_1-\frac{1}{t_1})-t_2}{2ik}\Delta\,,
\label{33}\\[2ex]
\left(\alpha-U_{22}\beta\right)\left(\alpha-U_{11}\beta\right)-\beta^2U_{12}
U_{21}&=&\frac{ik(t_1+\frac{1}{t_1})-t_2}{2ik}e^{-2ikc}\Delta \,.
\label{34}
\end{eqnarray}

Although the calculations that we shall introduce here in the sequel
are rather straightforward, their complexity makes it advisable to
give them with some detail. Otherwise the regular reader may have
unnecessary difficulties to reproduce the whole procedure.

Next, we write the matrix elements $U_{ij}$ in terms of $\Phi$ and
the $m_i$, for which we use (\ref{10}). We shall also use the
explicit form for $\alpha$ and $\beta$, which have been defined
after equation (\ref{18}). Then, (\ref{31}-\ref{34}) are transformed
into, respectively:

\begin{multline}\label{35}
-(2ck-1)^2+2(4c^2k^2-1) m_0 e^{i\Phi}-(2ck+1)^2 e^{2i\Phi}= \\[2ex]=-8  c
\frac{k(1+m^2b^2)+ima}{1-m^2b^2}e^{i \Phi}(m_2+im_1)e^{2ikc}\,,
\end{multline}

\begin{multline}\label{36}
(4c^2k^2-1) -2(4c^2k^2+1)m_0e^{i\Phi}-im_3 8cke^{i\Phi}+(4c^2k^2-1)
e^{2i\Phi}= \\[2ex]= -8 c \frac{2kmb+ima}{1-m^2b^2}e^{i \Phi}(m_2+im_1)\,,
\end{multline}

\begin{multline}\label{37}
-(4c^2k^2-1) +2(4c^2k^2+1)m_0e^{i\Phi}-im_3 8ck
e^{i\Phi}-(4c^2k^2-1) e^{2i\Phi}= \\[2ex]=-8 c
\frac{2kmb-ima}{1-m^2b^2}e^{i \Phi}(m_2+im_1)\,, \end{multline}

\begin{multline}\label{38}
(2ck+1)^2-2(4c^2k^2-1) m_0 e^{i\Phi}+(2ck-1)^2 e^{2i\Phi} =\\[2ex]=-8  c
\frac{k(1+m^2b^2)-ima}{1-m^2b}e^{i
\Phi}(m_2+im_1)e^{-2ikc}\,.\end{multline}

Then, we divide all these equations by $e^{i\Phi}$ and use
trigonometric relations to obtain:

%
%
%
\begin{eqnarray}
\label{39}\quad
(4c^2k^2+1)\cos{\Phi}+i4ck\sin{\Phi}-(4c^2k^2-1) m_0  
&=&4  c
\frac{k(1+m^2b^2)+ima}{1-m^2b^2}(m_2+im_1)e^{2ikc}\,, 
\\[2ex]
\label{40}
(4c^2k^2-1)\cos{\Phi}-(4c^2k^2+1)m_0-im_3 4ck &=& -4 c
\frac{2kmb+ima}{1-m^2b^2}(m_2+im_1)\,,
\\[2ex]
\label{41}
(4c^2k^2-1)\cos{\Phi} -(4c^2k^2+1)m_0+im_3 4ck  &=&4 c
\frac{2kmb-ima}{1-m^2b^2}(m_2+im_1)\,, 
\\[2ex]
\label{42}
(4c^2k^2+1)\cos{\Phi}-i4ck\sin{\Phi}-(4c^2k^2-1) m_0 &=&-4  c
\frac{k(1+m^2b^2)-ima}{1-m^2b^2}(m_2+im_1)e^{-2ikc}\,.
\end{eqnarray}
Then, subtract (\ref{40}) from (\ref{41}). It gives:

\begin{equation}\label{43}
im_3=\frac{2mb}{1-m^2b^2}(m_2+im_1)\,.
\end{equation}

Sum (\ref{40}) and (\ref{41}):

\begin{equation}\label{44}
(4c^2k^2-1)\cos {\Phi} - (4c^2k^2+1) m_0=-4c
\frac{ima}{1-m^2b^2}(m_2+im_1)\,.
\end{equation}

Sum (\ref{39}) and (\ref{42}):

\begin{equation}\label{45}
(4c^2k^2+1)\cos{\Phi}-(4c^2k^2-1)m_0=i8c
\frac{k(1+m^2b^2)\sin{2ck}+ma\cos{2ck}}{1-m^2b^2} (m_2+im_1)\,.
\end{equation}

Subtract (\ref{42}) from (\ref{39}):

\begin{equation}\label{46}
i4ck\sin{\Phi}=8c \frac{k(1+m^2b^2)\cos{2ck}-ma\sin{2ck}}{1-m^2b^2}
(m_2+im_1)\,.
\end{equation}

The system of transcendental equations equations (\ref{43}-\ref{46})
should give us the values of the parameters $\Phi$ and $m_i$ in
terms of $a$ and $b$. It is important to note that these equations
are complex as they have  real and  imaginary parts. Therefore, each
one splits into two equations, one corresponding to the identity of
its real parts and the other to the imaginary part. On the other
hand, we look for bound states, so that the solutions in $k$ must be
real. Then, the final result is a system of eight equations given
by:

\begin{eqnarray}
\label{47}
&\frac{2mb}{1-m^2b^2}m_2=0\,,
\\[2ex]
\label{48}
&\frac{2mb}{1-m^2b^2}m_1=m_3\,,
\\[2ex]
\label{49}
&4c \frac{ma}{1-m^2b^2}m_2=0\,,
\\[2ex]
\label{50}
&(4c^2k^2-1)\cos{\Phi}-(4c^2k^2+1)m_0 =4c\frac{ma}{1-m^2b^2}m_1\,,
\\[2ex]
\label{51}
&(4c^2k^2+1)\cos{\Phi}-(4c^2k^2-1)m_0=-8c
\frac{k(1+m^2b^2)\sin{2ck}+ma\cos{2ck}}{1-m^2b^2}m_1\,,
\\[2ex]\label{52}
&8c \frac{k(1+m^2b^2)\sin{2ck}+ma\cos{2ck}}{1-m^2b^2}m_2=0\,,
\\[2ex]\label{53}
&8c \frac{k(1+m^2b^2)\cos{2ck}-ma\sin{2ck}}{1-m^2b^2}m_2=0\,,
\\[2ex]\label{54}
&4ck\sin{\Phi}=8c
\frac{k(1+m^2b^2)\cos{2ck}-ma\sin{2ck}}{1-m^2b^2}m_1\,.
\end{eqnarray}

Since we are looking for a relation between the two independent
parameters $a$ and $b$ with $\Phi$ and the $m_i$,  these equations
cannot be independent. This systems looks to be hopeless, but it can
be solved with a little effort. Let us see how. First of all, it is
obvious that (\ref{47}) and (\ref{49}) give

\begin{equation}\label{55}
m_2=0\,.
\end{equation}

From equations (\ref{50}) and (\ref{51}), we manage the elimination
of $m_0$. If we multiply (\ref{51}) by $4ck^2+1$, (\ref{50}) by
$4ck^2-1$ subtract and divide by $4ck$, we obtain:

\begin{equation}\label{56}
4ck\cos{\Phi}
=-\frac{(4c^2k^2-1)ma+2(4c^2k^2+1)[k(1+m^2b^2)\sin{2ck}+ma\cos{2ck}]}{k(1-m^2b^2)}\;
m_1\,.
\end{equation}

Now, take (\ref{56}) and (\ref{54}), find their squares and sum. We
obtain an expression from where it is simple to write $m_1$ in terms
of $a$, $b$ and $k$. This gives:

\begin{equation}\label{57}
m_1=\frac{4ck(1-m^2b^2)}{\sqrt A}\,,
\end{equation}
with

\begin{multline}\label{58}
A=16c^2 \big[k(1+m^2b^2)\cos{2ck}-ma\sin{2ck}\big]^2+\\[2ex]+
 \frac{1}{k^2}\bigg[(4c^2k^2-1)ma+2(4c^2k^2+1)\big[k(1+m^2b^2) \sin{2ck}+ma\cos{2ck}\big]\bigg]^2\,.
\end{multline}

Once we have obtained $m_1$, we can get the value of $m_3$ through
(\ref{52}). Also,  dividing (\ref{54}) and (\ref{56}), we find:

\begin{equation}\label{59}
\tan{\Phi}=-8ck\frac{k(1+m^2b^2)\cos{2ck}-ma\sin{2ck}}{(4c^2k^2-1)ma+
2(4c^2k^2+1)[k(1+m^2b^2)\sin{2ck}+ma\cos{2ck}]}\,.
\end{equation}

It is noteworthy to say that, as we have eliminated $m_0$ from
(\ref{50}) and (\ref{51}), we could also have eliminated $\Phi$. We
can do it by multiplying (\ref{51}) by $4ck^2-1$ subtracting the
result of multiplying (\ref{50}) by $4ck^2+1$ and then dividing this
result by $4ck$.  We obtain:

\begin{equation}\label{60}
4ckm_0=-\frac{(4c^2k^2+1)ma+2(4c^2k^2-1)[k(1+m^2b^2)\sin{2ck}+ma\cos{2ck}]}{k(1-m^2b^2)}m_1\,,
\end{equation}
thus relating $m_0$ to $m_1$. As we have already commented,
relations $m_i$ are not independent but fulfil the relation
$m_0^2+m_1^2+m_2^2+m_3^2=1$. If we write $m_0$ and $m_3$ in terms of
$m_1$, we obtain:


\begin{equation}\label{61}
\left(\left[\frac{(4c^2k^2+1)ma+2(4c^2k^2-1)[k(1+m^2b^2)
\sin{2ck}+ma\cos{2ck}]}{4ck^2(1-m^2b^2)}\right]^2\right.
+\left. \left[\frac{2mb}{1-m^2b^2}\right]^2+1\right)m_1^2=1\,.
\end{equation}

Next, we use (\ref{57}) in (\ref{61}). After some manipulations, we
obtain a simple transcendent equation for $k$:

\begin{equation}\label{62}
k(1+m^2b^2)\sin{2ck}+ma\cos{2ck}=0\,.
\end{equation}

This equation can give us the energy values for given determinations
of the parameters $a$ and $b$. The variation of each of the first
three energy levels with $a$ and $b$ for $m$ fixed is given in the
Figure 1. Note that the parameters  $b$, $c$ and $m$ are always
positive and we have taken $a$ positive. The use of the Mathematica
tool called {\it manipulate} can give us the energy levels for
different values of $a$, $b$ and $m$. In Figure 1, we have chosen
the values given for the parameters, although the figure is quite
similar for another choices.

\begin{figure}
\includegraphics[width=0.75\textwidth]{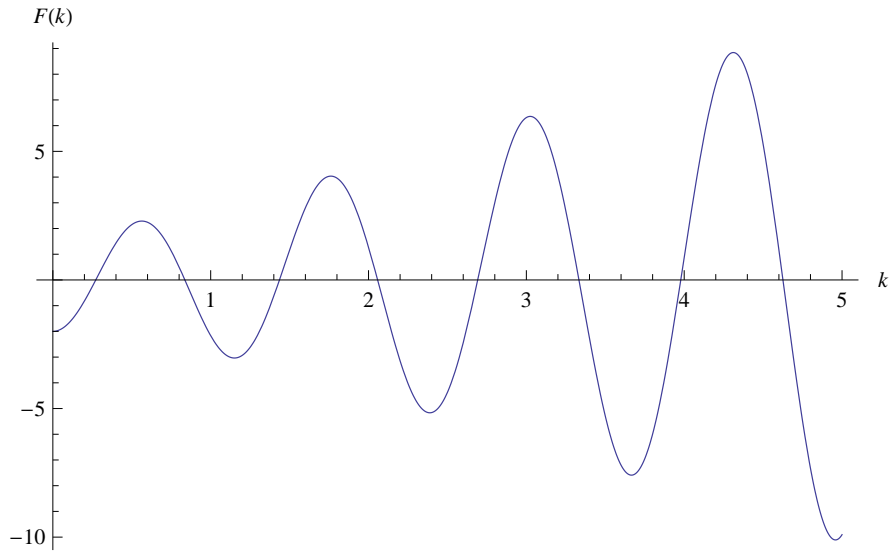}
\caption{First three energy levels with $a=4$, $b=2$, $c=2.5$ and
$m=0.5$.} \label{fig:1}
\end{figure}

The use of (\ref{62}) greatly simplifies some of the above
expressions. Now, we can write the parameters $m_1$ and $\Phi$ in
terms of $a$ and $b$:

\begin{equation}\label{63}
m_1=\frac{4ck(1-m^2b^2)}{ \sqrt{16c^2
[k(1+m^2b^2)\cos{2ck}-ma\sin{2ck}]^2+\frac{1}{k^2}[(4c^2k^2-1)ma]^2}}
\end{equation}
and

\begin{equation}\label{64}
\tan{\Phi}=-8ck\frac{k(1+m^2b^2)\cos{2ck}-ma\sin{2ck}}{(4c^2k^2-1)ma}\,.
\end{equation}

Then, we have to analyze formulas (\ref{63}) and (\ref{64}). One
would have expected that the relation between the parameters $m_i$
and $\Phi$ be one to one. Then, take one self adjoint extension of
$K$ characterized by the values of $a$ and $b$, i.e., take specific
values of these parameters in (\ref{1}). We would have expected that
these values give a unique pair of numbers for $m_1$ and $\Phi$.
However, (\ref{63}) and (\ref{64}) depend also on $k$ and therefore
on the energy levels.

The conclusion is that the relation between parameters is not one to
one, contrarily to what we may have expected.

We have an apparent difficulty with formula (\ref{64}). If we use
(\ref{62}) in (\ref{64}), $ma$ is simplified and we have an
expression like:

\begin{equation}\label{65}
\tan \Phi= \frac{8ck}{\sin 2ck}\;\frac{\cos 2ck-\sin^2
2ck}{4c^2k^2-1}\,.
\end{equation}

For $\Phi$ being fixed, this is a transcendental equation on $k$.
According to the inverse function theorem, one may at least locally,
obtain a relation of the form $k=h(\Phi)$. If we use this relation
in (\ref{63}), we finally obtain something like $m_1=F(\Phi,a,b)$,
which is not a desired relation.

Nevertheless, we have a cure for this problem and here is the correct treatment: From (\ref{62}) and using the
inverse function theorem, we can obtain local  expressions of the
type $k=\psi_n(a,b)$. We can use this  in (\ref{63}) and (\ref{64})
so as to obtain local relations of the type:

\begin{equation}\label{66}
m_1=F_n(a,b)\,,\qquad \Phi=G_n(a,b)\,.
\end{equation}

This result is somehow unexpected as it shows that the relation
between two different parameterizations of the self adjoint
extensions of the kinetic operator on the infinite square well is
not given by a unique function, but instead by a sequence of
functions depending on the energy levels. This means that for each
energy level, there is a distinct function that relates the values
of $a$ and $b$ with those of $m_1$ and $\Phi$ giving the same self
adjoint extension and therefore the same set of energy values.

\section{Concluding remarks}
\label{sec:4}

Being giving two specific values of $a$ and $b$ in (\ref{1}), the
number of energy levels for the Hamiltonian $H$ is infinite. This is
a fact shared by any self adjoint extension of $K$ \cite{BFV}.
Numerical estimations show that the largest deviations of the values
for the energy values given $E_n=k_n^2/2$, $k_n=n\pi/2c$ happens for
the lowest levels, being negligible for high values of $n$
\cite{GGN}. Then, for any value of $a$ and $b$, we give an infinite
series of values for $k$, say $k_n$. For any value of $k_n$, the
function that relates $a$ and $b$ to $m_1$ and $\Phi$ is different.
The somehow surprising conclusion of the present paper is that the
relations between different parameterizations of the self adjoint
extensions of $K$ are not simple as they are not given by a unique
equation as stated in the last section.

\section*{Acknowledgements}
Financial support is acknowledged to the Ministry of Economy and
Innovation of Spain through the Grant MTM2009-10751.



\end{document}